\documentstyle[aaspp4]{article}

\tightenlines

\begin{document}

\title{High Velocity Rain: The Terminal Velocity Model of Galactic Infall}

\author{Robert A. Benjamin } 
\affil{Department of Astronomy, University of Minnesota, Minneapolis, MN 55455\altaffilmark{1};
benjamin@astro.spa.umn.edu}

\author{Laura Danly}
\affil{Department of Physics and Astronomy, Pomona College, Claremont, CA 91711
\altaffilmark{2}; \\ ldanly@pomona.edu}

\altaffiltext{1}{also University of Minnesota Supercomputer Institute, 1200 Washington Ave South, Minneapolis, MN 55415}

\altaffiltext{2}{also Space Telescope Science Institute, 3700 San Martin Dr., Baltimore, MD 21218}

\begin{abstract}

A model is proposed for determining the distances to falling
interstellar clouds in the galactic halo by measuring the cloud
velocity and column density and assuming a model for the vertical
density distribution of the Galactic interstellar medium. It is shown
that falling clouds with $N(H~I) ~^{<}_{\sim}10^{19}~{\rm cm^{-2}}$
may be decelerated to a terminal velocity which increases with
increasing height above the Galactic plane. This terminal velocity
model correctly predicts the distance to high velocity cloud Complex M
and several other interstellar structures of previously determined
distance. It is demonstrated how interstellar absorption spectra alone
may be used to predict the distances of the clouds producing the
absorption.

If the distance, velocities, and column densities of enough
interstellar clouds are known independently, the procedure can be
reversed, and the terminal velocity model can be used to estimate the
vertical density structure (both the mean density and the porosity) of
the interstellar medium. Using the data of Danly and assuming a drag
coefficient of $C_{D} \cong 1$, the derived density distribution is
consistent with the expected density distribution of the warm ionized
medium, characterized by Reynolds. There is also evidence that for $z
~^{>}_{\sim}~0.4~{\rm kpc}$ one or more of the following occurs: (1)
the neutral fraction of the cloud decreases to $\sim$ $31 \pm 14\%$,
(2) the density drops off faster than characterized by Reynolds, or
(3) there is a systematic decrease in $C_{D}$ with increasing
$z$. Current data do not place strong constraints on the porosity of
the interstellar medium.

\end{abstract}


\section{The ``Hydrological Cycle'' of the Interstellar Medium}

``What goes up, must come down.'' This is  equally true for
the interstellar medium of the Galaxy as for the Earth, assuming that
the material in question does not have escape velocity.  Due to the
energy input in the midplane of the Galactic interstellar medium (ISM)
from supernovae (SN) and stellar winds, we know that matter goes up
(c.f, Spitzer 1990; McKee 1993).  But how does it come down? In this paper,
it is demonstrated  that there is potentially important dynamical information
contained in gaseous Galactic infall. Understanding the whole cycle is
important since, barring merger events, the global density and
pressure structure of the ISM is what determines the global star
formation and is therefore important in understanding
the evolution of the Galaxy.

First, what goes up? The mechanical energy input to the ISM is
dominated by stellar winds and supernovae. In the 1970's it was
realized that if this energy were converted into thermal energy much
of the volume of the ISM would be converted into hot, $T=10^{6}~K$ gas
(Cox \& Smith 1974; McKee \& Ostriker 1977). Since this gas is buoyant
and possibly overpressured, it will move away from the Galactic
midplane (Shapiro \& Field 1976; Norman \& Ikeuchi 1989; McLow \&
McCray 1988 ). However, these flows are subject to several
classes of instabilities (c.f., Field 1965; McLow \& McCray 1988;
Balbus 1988; Ferrara \& Einaudi 1992) Therefore, it is reasonable that
outflows, no matter how they occur in detail, are likely to form
denser, colder lumps.

In the ``galactic fountain'' model, these lumps are identified with
observed clouds of neutral hydrogen in the gaseous halo of the Galaxy,
clouds which on average are moving with negative velocities toward the
Galactic plane. Most notable in this regard are the so-called ``High
Velocity Clouds'' (HVCs) with $|v|> 70~ {\rm km~s^{-1}}$. These
clouds, first discovered by Muller, Oort, \& Raimond (1963), have
eluded all but one attempt to measure their distances (Danly, Albert,
\& Kuntz 1993), and are generally assumed to be located at greater
than a few kiloparsecs. The origin of these clouds and their lower
velocity companions is the subject of much debate and speculation
(c.f., Ferrara \& Field 1994, Wolfire et al. 1995). Regardless of
their origin, one can ask the following question: given the existence
of these neutral clouds, what are their dynamics? In particular, this
paper address the observational puzzle raised by Danly (1989) whose
ultraviolet absorption line study indicates that clouds appear to
decelerate rather than accelerate as they approach the Galactic disk.

Previous works on this subject fall into two classes, which although
illustrative, are incomplete in one important respect. The first class
of models attempted to treat the global characteristics of this infall
(Bregman 1980; Kaelble, de Boer, \& Grewing 1985; Wakker 1990; Li \&
Ikeuchi 1992). Since the sizes of the falling clouds are significantly
smaller than the Galaxy, the clouds were unresolved and it was assumed
that the clouds fell ballistically. This assumption allowed cloud
trajectories to be calculated for a large population of clouds, and
the resulting distribution of velocities with respect to direction on
the sky were calculated for comparison with the observational data. As
will be shown, the ballistic approximation is likely to be
inappropriate.

The second class of models calculates the full hydrodynamical
evolution of individual clouds (Tenorio-Tagle et al. 1987; Comeron \&
Torra 1992; Lepine \& Duvert 1994; Rand \& Stone 1996). These models
demonstrated how the impact of a high-velocity cloud of neutral
hydrogen onto the H~I disk could produce large-scale structures of H~I
and serve as a catalyst for star-formation. However, these models only
considered high column density
clouds, and neglected to take into account the fact that the Galaxy (or
at least the local solar neighborhood) is now known to have an
extended layer of ionized hydrogen, extending upward with a scale
height of 1 kpc and comprising 25\% of the total surface mass density
of the interstellar gas (Reynolds 1993).

This paper discusses some simple considerations on the nature of
Galactic infall and shows how the study of its dynamics has the
potential to greatly improve our knowledge of the structure of the
ISM. It is argued that drag forces on falling interstellar clouds are
sufficient to lock many clouds to a terminal velocity which depends on
the cloud height and column density.  Measurement of a cloud velocity
and column density may then be converted to
$z$-height.  Preliminary versions of this work have been presented as
conference proceedings (Benjamin 1996a,b). In \S 2, the basic model is
presented and it is argued that many clouds may be
falling at a terminal velocity. \S 3 presents some
potential applications of this concept to observational data. Finally
\S 4 lists the principal results and implications of this work
and notes the similarities between the proposed model and a more
familiar but analagous process, namely the dynamics of terrestrial
rain.

\section{The Dynamics of Infall}

Neutral hydrogen clouds above the Galactic plane must be surrounded by
an external (ionized) gaseous medium, or they would disperse on a
short timescale (Spitzer 1956). What is the effect of this gaseous
medium on a cloud's dynamics? Since neutral clouds are less buoyant
than their surroundings, they will fall. For simplicity, it is assumed the cloud is
restricted to move in the $z$ direction. Assuming a simple $v^{2}$ dependence
for the drag force, the equation of motion is 

\begin{equation}
m_{c}\frac{dv_{c}}{dt}=\frac{1}{2}C_{D}\rho_{h}(z)[v_{c}-v_{h}(z)]^{2}A_{c}-
m_{c}g(z)~.
\end{equation} \label{eq-momeq}

The mass, surface area, and velocity of the cloud are 
$m_{c}$, $A_{c}$, and $v_{c}$; the halo gas density, velocity, and
gravitational acceleration at height $z$ above the Galactic plane are
$\rho_{h}(z)$, $v_{h}(z)$, and $g(z)$. The drag coefficient, $C_{D}$,
indicates the efficiency of momentum transfer to the cloud, the
accompanying factor of $1/2$ is conventionally (but unfortunately not
always) included. The first term in the right hand side represents the
decelerating force of ram pressure; the second is the accelerating
force of gravity. If these two forces are in balance, $dv/dt=0$ and
the cloud has reached ``terminal velocity'', $v_{T}$. Ballistic
calculations mentioned in the previous section ignored the
deceleration term. Hydrodynamical calculation of high velocity clouds
hitting the disk underestimated the deceleration by underestimating
$\rho_{h}(z)$ or used a high $m_{c}/A_{c}$.

It is assumed here that (1) the cloud can maintain itself as a
discrete entity, and (2) the above equation provides a first
approximation to its motion.  The first point is at issue because
before the cloud approaches terminal velocity it is subject to
Rayleigh-Taylor instabilities, and it is always subject to
Kelvin-Helmholtz instabilities. Recent numerical simulations (MacLow
et al. 1994; Jones, Ryu, \& Tregillis 1997) have demonstrated how
inclusion of even weak magnetic fields may inhibit these
instabilities. The second point is at issue because although it clear
that the drag force must increase with velocity, it is not known what
functional form will best characterize the drag. Equation (1) assumes
that is the pressure difference from the front to back of the cloud,
i.e. the ram pressure, which is responsible for the deceleration. The
retarding force is thus proportional to $v^{2}$ with constant of
proportionality $C_{D}$. In actuality, the process of deceleration is
probably more complicated and depends upon the viscosity of the ISM,
the deformation of the cloud, and the flow pattern around the
cloud. This issue is best resolved through numerical simulations.  The
most recent simulations of Jones et al (1997) indicate that the global
deceleration of a cloud is proportional to $v^{2}$, and $C_{D} \cong
1$, although it was necessary to modify this drag coefficient to take
into accout the lateral expansion of the cloud and the presence of
magnetic fields.

\subsection{The Terms and Their Uncertainties}
We now discuss each of the terms in equation (1), the values used for this
paper and their  uncertainties:

$g(z)$: The gravitational acceleration is taken from Wolfire et al.
(1995) and Spergel (1994). This is shown in Figure 1, and
can be fit to within 15\% by $g(z)=9.5 \times 10^{-9} \tanh
(z/400~pc)$. At some large height, the assumption of purely vertical
trajectory will break down, the value of this depends on the initial
conditions for the cloud, e.g., whether it starts out initially
co-rotating with the disk below, etc.

$m_{c}$: The mass of the cloud may either increase or decrease over
time. Which occurs and to what degree depends upon the particular
environmental conditions and assumed input physics. The cloud can gain
mass by ``sweeping up'' either ambient halo gas or other slower moving
clouds as it falls. On the other hand, Kelvin-Helmholtz and possibly
Rayleigh-Taylor instabilities will act to strip the cloud of mass and
could potentially disrupt the cloud entirely (Klein, McKee, \& Colella
1994; Jones, Kang, \& Tregillis 1994). Inclusion of modest magnetic
fields will ameliorate this effect (MacLow et
al. 1994; Jones, Ryu, \& Tregillis 1997). Without detailed
simulations, it is hard to know what the net effect is on the cloud
mass and motion. Here a constant mass is assumed.

$A_{c}$: It is assumed that the cloud is flattened in the direction
of motion. The cloud mass is $m_{c}=\mu A_{c}\int n_{c}(z)~ dz=\mu
A_{c} N_{c}$, where $n_{c}(z)$ is the cloud particle density, $N_{c}$
is the column density, and $\mu$ is the mean mass per particle. The
surface area term in the equation of motion may then be divided
out. The geometry of the cloud, i.e.  whether it is rod-like (high
$m_{c}/A_{c}$) or sheet-like (low $m_{c}/A_{c}$), will affect the
amount of drag. By making this assumption, the effects of the
geometry, and the associated uncertainties, have been subsumed into
the drag coefficient, $C_{D}$.

$C_{D}$: The uncertainty in this coefficient will need to be addressed
by numerical simulations. For a perfectly streamlined cloud,
$C_{D}=0$; for perfect momentum transfer, $C_{D}=2$. $C_{D}$ can
exceed 2 if the flow pattern yields an underpressurized region behind
the cloud. This work uses $C_{D}=1$ as a starting guess (Jones, Ryu, \&
Tregillis 1997). But $C_{D}$ may depend on several factors: the mass
and velocity of the cloud, the halo density and temperature, the
relative importance of magnetic fields, radiative cooling, etc. And it
may depend upon the history of the cloud, i.e. the drag coefficient will
change over time as the cloud deforms and tries to adjust to its
motion through the background. Comparison of indvidual cloud
morphologies (Odenwald 1988; Wakker \& Schwarz 1991) to simulations of
cloud-halo interactions will provide independent constraints on the
relevant input physics and the proper value of the $C_{D}$.

$n_{h}(z)$: The density structure of the gaseous halo is the great
unknown. If the halo were uniform, isothermal, static, and dominated
by thermal pressure, a simple application of the equation of
hydrostatic equilibrium would yield the density structure (Spitzer
1956; Wolfire et al. 1995).  However, this situation is
far from being the case, with nonthermal pressure being more important
than thermal pressure sources at least up to a few ${\rm kpc}$ (Boulares \&
Cox 1990). With many of the theoretical issues relating to the density
structure of the interstellar medium still outstanding, it is probably
far safer to stick to using the observationally determined density
structure. Three different density distributions are considered and
shown in Figure 1:

(A) The mean density of the warm ionized layer of H II using the
parameterization of Reynolds (1993): $n_{h}(z)=0.025~e^{-z/910~pc}~cm^{-3}$.

(B) The ``Reynold's layer'' plus the mean H I density of Dickey \&
Lockman (1990), which consists of three components: two Gaussians of
central densities $0.395$ and $0.107~cm^{-3}$ and FWHM of $212$ and
$530~pc$, and an exponential with central density $0.064~cm^{-3}$ and
scale height of $403~pc$.

(C) The above two plus an isothermal ``hot'' halo with $T=10^{6}~K$ as
prescribed by Wolfire et al. (1995), $n_{h}(z)=1.1 \times 10^{-3} 
(1+(z_{{\rm kpc}}^{2}/19.6))^{-1.35}~cm^{-3}$.

The third component, unlike the first two is not an observationally
derived quantity, but is constructed by assuming the presence of a
hydrostatic isothermal $T=10^{6}~K$, halo, and determining the
midplane density by matching the X-ray emission data of Garmire et
al. (1992). The only independent information on the halo gas
density above $\sim 3~{\rm kpc}$ is by studying neutral cloud
morphologies and dynamics. For instance, the diffuse $H \alpha$
observations of a neutral Magellanic Stream cloud by Weiner \&
Williams (1996) together with assumptions on the nature of the
cloud-halo interaction indicate a gas density for the halo of
$n_{h}=10^{-4}~cm^{-3}$ at $z >> 10~{\rm kpc}$.

$v_{h}(z)$: It is assumed that the velocity of the background medium
is static with respect to the Local Standard of Rest. This assumption
may break down close to the Galactic plane, where SN and stellar wind
driven turbulence drive large scale mass motions.  It must also break
down well above the plane, since it is not possible that the halo
corotates with the disk up to infinity. At some height, the halo must
lag behind the disk. This tendency is possibly observed by Rand
(1997), whose observations of NGC 891 indicate that at a height of
$5~{\rm kpc}$, the $H \alpha$ rotation curve lags rotation in the
plane by approximately $15~ {\rm km~s^{-1}}$.

\subsection{The Terminal Velocity Model}

In the simplest of all possible galaxies, interstellar clouds would all be
falling at their terminal speed,

\begin{equation}
|v_{T}(z)|=\sqrt{\frac{2g(z)N_{H~I}}{C_{D}f_{c}n_{h}(z)}}~,
\end{equation} \label{eq-vt}

where the cloud neutral fraction is
$f_{c}=N_{H~I}/(N_{H~I}+N_{H~II})$.  This factor is introduced since
what matters in cloud dynamics is the {\it total} column density but
observational techniques generally only allow measurment of the
neutral column density. In fiducial units, and using the analytical
formulation for the density structure of the Reynold's layer, this is

\begin{equation}
|v_{T}(z)|=26~{\rm km~s^{-1}}(C_{D}f_{c})^{-1/2}\left(\frac{N_{H~I}}{10^{19}~cm^{-
2}}\right)^{1/2} \left(\frac{g(z)}{g(1~{\rm kpc})}\right)^{1/2}
\left(\frac{n_{h,o}}{0.025~cm^{-3}}\right)^{-1/2} e^{z/2H}
\end{equation} \label{eq-vtfid}~.

Since typical halo cloud speeds are similar to the value of $|v_{T}|$,
the effects of drag can not be automatically dismissed. The appendix
shows the typical time and distances necessary for a cloud with
constant gravitational acceleration to converge to terminal velocity
in both a uniform and a porous interstellar medium. Since halo density
and gravitational acceleration change as a function of position, the
terminal speed is position dependent.  $|v_{T}(z)|$ is defined as the
local terminal speed; this is plotted for clouds of varying column
density and different halo density structures in Figure 2.

Do clouds reach a terminal speed? To answer this question, equation
(1) is used to calculate the trajectories of clouds with
$N_{H~I}=10^{18}~{\rm cm^{-2}}$ , $10^{19}~{\rm cm^{-2}}$ , and
$10^{20}~{\rm cm^{-2}}$ dropped at rest from heights of
$z_{i}=1,5,~{\rm and} ~8~{\rm kpc}$. The trajectories, together with
the ballistic trajectories and the local terminal velocity curves, are
shown in Figure 3 for density structure of the warm ionized medium
only. The points note the location after $n$ ($n=1,2,..$) free-fall
times, $t_{ff}=\sqrt{2z_{i}/g(z_{i})}$. The highest velocity that can
be achieved by a falling cloud is the escape velocity, $v_{esc}
\approx 500~ {\rm km~s^{-1}}$ (Binney \& Tremaine 1987).

For clouds of column density $N(H~I)=10^{20}~{\rm cm^{-2}}$, the cloud
is significantly decelerated from a ballistic trajectory, but its
velocity never closely resembles the terminal velocity.  On the other
hand, these clouds are relatively rare. For reference, the total H I
column density through the H~I disk is $6 \times 10^{20}~{\rm
cm^{-2}}$ (Dickey \& Lockman 1990).  Clouds with $N(H~I)=10^{19}~{\rm
cm^{-2}}$ are much more significantly decelerated, and within 1 kpc of
the disk, their velocities tend to be within 50\% of the local
terminal velocity.  Most usefully, clouds with $N(H~I)=10^{18}~{\rm
cm^{-2}}$ will rapidly lock on to the local terminal velocity. These
clouds spend a proportionally longer period of their lifetime at or
near terminal velocity rather than in the ballistic phase. The rest of
this paper will use the shorthand description ``low'', ``medium'' and
``high'' column density for clouds of $N(H~I)=10^{18}~{\rm
cm^{-2}},10^{19}~{\rm cm^{-2}},~ {\rm and}~10^{20}~{\rm cm^{-2}}$,
respectively.

These calculations indicate that the assumption of terminal velocity
is reasonable starting point for predicting the cloud velocities for
clouds with $N(H I) ~^{<}_{\sim} 10^{19}~{\rm cm^{-2}}$, with
increasing accuracy for decreasing cloud column density.

\section{The Terminal Velocity Model: Tests and Applications}

The terminal velocity model allows a distance prediction for every
interstellar cloud with a negative vertical velocity and measured
column density. Although the range of applicability of the model must
ultimately be tested by observational studies, it is clear that there
exist certain fundamental limits on this method. First, given the
existence of postive velocity clouds, it is clear that equation (1)
must be modified to include additional sources of momentum. It is
reasonable to assume that the same momentum sources will also affect
some subset of downward travelling clouds. Second, as shown above,
clouds will have some sort of dynamical history, and this history may
affect the space velocity. At one extreme, the space motion of a high
column density cloud in low density enviroments, such as a cloud in
the Magellanic Stream, will largely retain a ``dynamical memory'' of
its origin. At the other extreme, a cloud which forms {\it in situ} in
the halo may slowly grow in column density, always adjusting its
velocity to match the local terminal velocity, thus retaining no
``memory'' of its previous motion. 

There are three primary ways in which this model may be
applied to observations:

1) Assume $C_{D}$ and $n_{h}(z)$.  Measurement of $N_{H~I}$ and $v$
allows determination of the distance, $z$.  

2) Assume $C_{D}$. Measurement of $N_{H~I}$, $v$, and $z$ allows 
determination of the density structure of the halo, $n_{h}(z)$.

3) Assume $n_{h}(z)$. Measurement of $N_{H~I}$, $v$, and $z$ allows
an empircal calculation of the drag coefficient, $C_{D}$.

It seems likely that the first application will be the most useful,
\S~\ref{sec-distpred} shows how the model may be invaluable for
interpreting UV and optical absorption line data. The second
application in \S~\ref{sec-dstruct} will require a large and
systematic effort of UV and optical absorption line studies to
independently obtain cloud distances, and could provide an invaluable
probe for the mean density and porosity of the
interstellar medium. The third application, also discussed in
\S~\ref{sec-dstruct} will probably only become necessary if
theoretical considerations are not able to sufficiently limit the
range of variation of $C_{D}$.

\subsection{How to Predict Cloud Distances} \label{sec-distpred}

\subsubsection{Tests}

Figure 4 shows negative velocity clouds with $|b|>30$ and $|z|
^{>}_{\sim} 200~pc$ for which distance limits have been obtained. Data
are taken from Kuntz and Danly (1996), Danly et al. (1992), Albert et
al (1993), de Boer et al. (1994), Benjamin et al. (1996), Wesselius \&
Fejes (1973), and Benjamin, Hiltgen, \& Sneden (1997).  Also included
is a line indicating the negative velocity extent of UV absorption as
a function of star height from Danly (1989). This last study was able
to put onto a more quantitative footing the already empirically
recognized fact that the higher the cloud velocity, the more distant
it is likely to be (c.f., Welsh, Craig, \& Roberts 1996; de Boer et
al. 1994). This figure shows, despite a paucity of accurate distances,
a trend of velocity increasing with $z$. Here, the model is compared
against clouds of known distance, namely Complex M (Danly, Albert, \&
Kuntz 1993), the ``IV Arch''(Kuntz \& Danly 1996), the Ursa Major IVC
(Benjamin et al.  1996), and several clouds observed by Albert et al
(1993). The clouds chosen have $|b|>50$, in order to minimize the
effects of galactic rotation, and the values of velocity and column
density used in equation (2) are corrected for projection
effects assuming $v_{z}=v|csc~b|$ and $N_{z}=N_{H~I}|sin~b|$.  The
observed cloud parameters and their references are shown in
Table~\ref{tbl-predict1}, along with a comparison between the distance
predictions and the observed distance limits. In all cases, it is
assumed that $C_{D}=1$ and $f_{c}=1$, except for the IV Arch, where
independent information indicates that $f_{c} \cong 0.44$ (see
discussion in the next section). Predictions are shown for density
models A and B.

Allowing for uncertainties in the stellar distances (typically $25
\%$), the model successfully predicts the distances for eight of ten
clouds, including the HVC Complex M, assuming the warm ionized medium
is principally responsible for their deceleration. For clouds at
sufficiently high $z$, these distances predictions probably tend to be
overestimates since the ambient pressure is lower, and the expected
neutral fraction of clouds is expected to be $f_{c}<1$ (Songaila,
Cowie, \& Weaver 1988; Ferrara \& Field 1994). Accounting for this
will have the effect of reducing the predicted distance by $\Delta z
\cong H\ln(f_{c}^{-1})$, where $H=910~pc$ for the warm ionized density
distribution.

The case of the Ursa Major cloud deserves special comment. The model
correctly predicts this distance to this high column density molecular
cloud, assuming that both the H I and H II layer
contribute to its deceleration. Since this cloud has sufficiently high
column density that it should be travelling faster than much of the H
I that makes up the high latitude H I layer, it is not surprising that
it should feel the decelerating force of the H I layer. Since the
other H~I clouds are presumably also falling, the ram pressure due to the H~I
layer will be reduced because the relative velocity of the high column
density cloud of interest and the ``typical'' H~I cloud. If 
$R=N_{c}/N_{l}$ is the ratio of the column density of the heavy cloud
to the column density of the ``typical'' cloud making up the H I
layer, and $n_{H~I}$ and $n_{H~II}$ are the mean densities of the H
I and H II layer, then the terminal velocity of the heavy cloud,
$v_{c}$ is

\begin{equation}
v_{c}=v_{l}\left[ \frac{n_{H~I}+
\sqrt{Rn_{H~II}^{2}+(R-1)n_{H~I}n_{H~II}}}{n_{H~I}+n_{H~II}} \right]
\end{equation}

where $v_{l}=\sqrt{2N_{l}g/(C_{D}n_{H~II})}$ is the terminal velocity
of the ``typical'' H~I layer cloud in the H~II layer. Solving for the
velocity difference between the cloud and H~I layer that allows a 
match between the terminal velocity model and the observed distance yields
$v_{l}=23~{\rm km~s^{-1}}$, which corresponds to a cloud column density of
$N_{l}=10^{18.9}~{\rm cm^{-2}}$. 

\subsubsection{Predictions}

The line of sight toward HD 93521 ($z= 1.5 \pm 0.4~ {\rm kpc}$,
$b=62.2$) is one of the best studied halo lines of sight, and contains
an unusually large number of absorption components. It therefore
provides a good test case for predicting cloud distances. Here, the
terminal velocity model is used to predict the distances to the clouds
observed in UV absorption line data of Spitzer \& Fitzpatrick
(1993). The measured velocities, column densities, and predicted
distances are in Table~\ref{tbl-predict2}. The $N(H~I)$ column density
is estimated by taking advantage of the fact that S II is a very good
tracer for H I, i.e. the ratio N(S~II)/N(H~I) is relatively constant,
$N(H~I)_{s}=10^{4.76}N(S~II)$ (c.f., Spitzer \& Fitzpatrick
1993). The advantage of this assumption is that a single
self-contained measurement may be used to determine cloud
distance; both the cloud column density and velocity may be obtained
from a single spectrum. Other ions (including the optically
detectable ${\rm Ca^{+}}$ or ${\rm Ti^{+}}$) could also be used provided a
similar ``correction factor'', which allows conversion of $N_{ion}$ to
$N(H~I)$. For many ions, however, it has been demonstrated that this
``correction factor'' (which depends on depletion and ionization
effects) is velocity dependent, so care must be taken (c.f., Jenkins
1987).

The cloud neutral fraction, $f_{c}$, is determined by comparison of
the ratio of column densities of two ions to photoionization models of
the cloud. For the data of Spitzer \& Fitzpatrick (1993) the ratio
$N(S~III)/N(S~II)$ is used. Theoretical values for this ratio are
calculated using version 84.12a of the photoionization code CLOUDY
(Ferland 1993) to model the cloud as a plane-parallel, constant
density slab photoionized by the predominantly stellar ionizing
spectrum of Bregman \& Harrington (1986) or the X-ray emission
spectrum of Benjamin \& Shapiro (1996). (Note that given the cloud's
deceleration, the constant density assumption is unlikely to be
valid. The effect of a non-uniform density on the ionization structure
is deferred to future work.) For $-4.5 < \Gamma < -2$ and stopping H I
column densities $18 <logN(H~I)_{end}< 19.5$, the column density ratio
is well fit to within 20\% by

\begin{equation}
\log{\frac{N(S~III)}{N(S~II)}}=0.9 \log \Gamma + B-0.25N ~,
\end{equation}

where $N=logN(H~I)_{end}-18$ and $B=2.90$ for the Bregman \&
Harrington spectrum and $B=3.25$ for the Benjamin \& Shapiro
spectrum. For the Bregman \& Harrington ionizing spectrum, the
ionization parameter required to match the observed values lies in the
range $\log~\Gamma=-4.25$ to $-3.60$. $f_{c}$ can be estimated by
balancing the ionizing photon flux, $\phi$ (H ionizing photons
$s^{-1}$ ${\rm cm^{-2}}$), with the total number of recombinations in
some column of length, $l$, so that $\phi \cong n_{H}^{2} \alpha l$,
where $\alpha \cong 3 \times 10^{-13} {\rm cm^{3}}~s^{-1}$ is the
recombination coefficient. Since the ionization parameter is
$\Gamma=\phi/(c n_{H})$, $N(H~II)=\Gamma c/\alpha$, and therefore
$f_{c}=N_{H~I}/(N_{H~I}+(c \Gamma /\alpha))$. The resultant values of
$f_{c}$  are in Table 2. Use of the X-ray ionizing spectrum of Benjamin
\& Shapiro (1996) would increase $f_{c}$ by $\sim 0.2$.

One test of the accuracy of the predictions is that the derived cloud
distances must not exceed the distance of the star. Within the
estimated uncertainties, they satisfy this criterion, with the
exception of component 5, which is a marginal detection. High
resolution optical spectroscopy of suitable target stars in the
vicinity of HD 93521 could be used to test these predictions.  Albert
(1983) has already reported observations for 32 LMi ($z=150~pc$) a
star separated from HD 95321 by $3.8^{\circ}$.  She detects no
absorption at any velocity. This nondetection is consistent with all
the distance predictions, with the possible exception of component 7
at $v=-10.2~{\rm km~s^{-1}}$. However, for the lowest velocity
components, small errors in measurement of the velocity can lead to
large errors in the derived distance. This fact combined with the
likelihood that additional momentum sources will be important close to
the galactic plane means this method may not be a reliable way to
determine cloud distances below $\sim 200~ pc$. Nevertheless, this
type of analysis can be repeated wherever one has UV and optical
absorption line data and holds the promise of significantly improving
our knowledge of the structure of the interstellar medium.

\subsection{How to Determine the Density Structure of the Halo} \label{sec-dstruct}

If there existed a large population of interstellar clouds with known
distances, velocities, column densities, and neutral fractions, the
procedure described above could be inverted, and one could use these
clouds plus the terminal velocity model to constrain the mean density
structure of the gaseous halo, the porosity, and the variation with
galactic longitude. As Figure 4 indicates, the number of cloud for
which tight distance brackets have been obtained is regretably
small. However, one can use the UV absorption study of Danly (1989) to
address this question, assuming that the distance to the star is
approximately the distance to the cloud producing the most negative
velocity absorption. This assumption is more likely to be true in the
North Galactic Hemisphere, which appears to be full of infall at a
large range of $z$ and $v$ (Danly 1989; Kuntz \& Danly 1996) than in
the South Galactic Hemisphere, which has less infalling gas.

\subsubsection{The Mean Density of the Gaseous Halo}

Rearranging the terms in equation (2), the
mean density of the halo as a function of $z$ as derived by 
cloud kinematics is  

\begin{equation}
C_{D}f_{c}n^{kin}_{h}(z)=\frac{2 N_{H~I}g(z)}{v^{2}}
\end{equation} \label{eq-nh}

By measuring the column density and velocity of cloud and estimating
the gravitational acceleration at the location of the cloud, one can
solve for the local ambient density at the location of the cloud,
short of uncertainties in the appropriate value of the drag
coefficient and ionization fraction.  Table~\ref{tbl-halostruct} lists
some of the target stars of Danly (1989), their height above the
Galactic plane, $z_{star}$ (which is assumed to equal to $z_{cloud}$),
the most negative velocity of absorption, $v_{-}$. The sample has been
limited to lines of sight with $|b|>50$ in order to avoid
uncertainties associated with galactic rotation and projection
effects, and $N(H~I)>10^{17.9}~{\rm cm^{-2}}$ where the
column densities become highly uncertain.  For the cases of BD +38
2182 and HD 93521, the data have been updated to be consistent with
those in Table 1. The column density $N(H~I)$ in the velocity range
$v_{-} \pm 5~{\rm km~s^{-1}}$ density is obtained either by
integrating the H~I profiles (Danly et al 1992) in the direction of
each star or, when no profiles were available, by locating the stars
on the H~I emission maps of Kuntz \& Danly (1996). Uncertainties
were not calculated for the derived density, since they were not
available for all input parameters. 

For each line of sight, the kinetically derived  gas density times
$C_{D}f_{c}$ is tabulated. The derived density at low $z$ ($z<100~pc$)
is in good agreement with the midplane density of the warm
ionized medium (Weisberg, Rankin, \& Boriakoff 1987) although it
arises from an independent line of reasoning. Given the expectation
that the model should break down at low $z$ due to additional momentum
sources and that the clouds producing this absorption lie within the
expected confines of the Local Hot Bubble (Cox \& Reynolds 1987), this
agreement may merely be fortuitous.

For the more distant stars (and presumably more distant clouds) the
derived density decreases as a function of $z$, in agreement with
expectations. However, the derived density drops off faster than one
would expect in the density models considered. This is shown in
Table~\ref{tbl-halostruct}, where the ratio of the derived kinematic
density to the expected density, $n^{kin}(z)/n_{h}(z)=C_{D}f_{c}$ is
tabulated. Figure 5 illustrates this graphically for the warm ionized
medium density.  For clouds (stars) below 400 pc, $C_{D}f_{c}=0.92 \pm
0.09$, where the line of sight towards HD 219688 has been excluded as
anomalous, and the uncertainty is the standard deviation of the
derived values and does not reflect measurement errors in the input
parameters. For stars between 500 pc and and 3300 pc, $C_{D}f_{c}$
decreases to $0.31 \pm 0.14$. The data do not allow one to
characterize whether there is smooth or abrupt decrease. Above
3300 pc, $C_{D}f_{c}$ increases to values significantly greater than
one.

If $C_{D} \cong 1$, one explanation for this trend is that the cloud
neutral fractions decline to $f_{c} \approx 0.3$ for
$|z|~^{>}_{\sim}~400~{\rm pc}$. This trend is consistent with
theoretical expectations (Ferrara \& Field 1994), and the value is
consistent with the derived neutral fractions obtained by comparison
of column densities of S II and S III with photoionization model as
described in the previous section. Other possible explanation are a
systematic decrease of $C_{D}$ at high $z$, or the mean density
droping somewhat faster the exponential model used by Reynolds(1993).
Differentiating among these possibilities will require a more thorough
observational investigation, plus a better theoretical
characterization of how $f_{c}$ and $C_{D}$ might vary with height.

It is important to remember that equation (2) yields the halo density
at the location of the cloud, but the $z$ reported is that of the star
in which absorption is detected. Therefore the reported value of
$n^{kin}(z)$ will be an upper limit on the gas density at $z_{star}$,
since the gas density will decrease between $z_{cloud}$ and
$z_{star}$. Thus if clouds are confined to a layer with $z_{max}$,
$C_{D}f_{c}$ will be larger than its true value if $z_{star} \geq
z_{max}$. This provides a viable explanation for the large
increase in $C_{D}f_{c}$ for $|z| > 3300~pc$.

\subsubsection{The Porosity of the Halo}

Once the effect of column densities is factored out, assuming a model
for the mean density of the gaseous halo defines a position-velocity
curve.  If one assumes that regions of
hot, $T \cong 10^{6}~K$ gas in the halo are in pressure equilibrium
with the warm, $T=10^{4}~K$ ionized gas, the density in the hot
patches will be $\sim 100$ times lower. Every time a cloud hits one of
these patches, it will start to accelerate, introducing scatter and widening
the position-velocity curve into a band. This provides an upper limit on the
porosity, since observational error, uncertain projection effects, and
Galactic rotation will introduce their own scatter into the
velocity-distance relationship.

Unfortunately at the current time, the data barely define a good mean
value, let alone a scatter. In the appendix, it is shown how the
velocity range, $\Delta v$ depends on the linear filling fraction, $f$,
and ``cell length'', $L$, for ISM with constant mean density.  The
following is an example of how this may be used to estimate the effect
of patchiness on cloud dynamics: Assume that the velocity scatter in
the terminal velocity relationship for clouds of column density
$N(H~I)=10^{18}~{\rm cm^{-2}}$ is found to be $\Delta v = 10~ {\rm
km~s^{-1}}$ at all $z$.  Convert $\Delta v$ to $\Delta \tilde{v}(z)$
by dividing by the terminal velocity shown in Figure 2. These values
can be converted to a loci of filling factor and cell length using
Figure A1. Assuming a filling factor sets the normalized cell length,
which can be converted to a physical length by multiplying by the
convergence length $z_{T}=v_{T}(z)^{2}/g(z)$. If $f=0.8$, to introduce
scatter greater than $\Delta v=10~{\rm km~s^{-1}}$ would take cell
lengths of greater than $L_{cell}\sim~1~ {\rm kpc}$, much larger than
the scales over which the terminal velocity will change.  However, if
$f=0.2$, then cell lengths must only be greater than $L_{cell}\sim
200~pc$, a value low enough to be of some interest. Since, in general,
the cell lengths are sufficiently large that they approach the scale
on which the mean density and gravitational acceleration are changing,
a better method to estimate of the scatter around the terminal
velocity curve would be to perform Monte Carlo simulations of
clouds falling through different assumed density distributions. This
is deferred to later work.

It is important to note that the pattern of cloud velocity increasing
with height observed by Danly (1989) {\it already indicates that the
density of the halo ISM is at some level homogeneous}, at
least on the scales probed by galactic infall, unless one argues that
that the observed correlation is a chance occurence or there is some
alternative explanation. Danly's results set an upper limit on the
amount of scatter in the position-velocity curve. This corresponds to
$\Delta v ~^{<}_{\sim} 30~{\rm km~s^{-1}}$. Unfortunately, cell sizes
necessary to produce such scatter are sufficiently large ($z > 1~{\rm
kpc}$) as not to provide any interesting limits on the porosity of the
interstellar medium.

\section{High velocity rain}

There is a time-honored connection between the study of neutral halo
clouds and understanding of the gaseous halo of the Galaxy.  The {\it
existence} of these clouds discovered by Munch \& Zirin (1961)
prompted Spitzer (1956) to infer the {\it existence} of a large
gaseous extent for the Galaxy. Here, studies on the {\it dynamics} of
the same clouds by Danly (1989) and also by Albert (1983), have
prompted this work which is an attempt to characterize the {\it
structure} of the ambient interstellar gas at high $z$. The density
structure derived from cloud kinematical arguments, it is argued, is
similar to the density of the warm ionized medium as determined by
pulsar dispersion measures, and shows evidence for a decrease in 
cloud neutral fractions with increasing $|z|$.

The critical factor that differentiates this work from previous
characterizations of galactic infall is the identification of the
recently discovered extent of the warm ionized medium (Reynolds 1989)
as occupying the intercloud regions. The warm ionized medium,
comprising 25\% of the total surface mass density of the interstellar
medium, must have an important effect on the trajectories of neutral
clouds. Instead of treating the cloud motions as ballistic in a low
density, dragless intercloud medium (c.f., Oort 1954, Field \& Saslaw
1965, McKee \& Ostriker 1977), the terminal velocity model represents
the opposite extreme. Instead of {\it no} drag forces, cloud motions
are governed {\it entirely} by drag forces. Instead of cloud motions
being essentially random, they are systematic. Since the motions are
systematic, their distances can be predicted and tested. The model
correctly predicts the distances of eight of ten interstellar clouds
for which data were available. 

In characterizing the dynamics of Galactic infall, Danly (1989)
referred to falling clouds of low column density as ``drizzle'' and
clouds of high column density as ``storms''. The terminal velocity
model suggests that this analogy extends to the dynamical level. Much
of the physics discussed here applies equally well to the dynamics of
raindrops as to falling interstellar clouds (c.f. Foote \& du Toit
1969).  In both cases, there is a phase change, a loss of buoyancy,
and an acceleration up to a terminal velocity in a (presumably)
exponential density structure.  There are also important
differences, most notably that liquid water unlike neutral
hydrogen clouds is incompressible. Nevertheless, this has been a
useful analogy in developing this work, and questions that arise in
comparing the two systems may provide interesting avenues for future
exploration.

\acknowledgements

We would like to acknowledge several researchers for useful advice and
suggestions, most notably Don Cox, John Dickey, Robbie Dohm-Palmer,
Andrea Ferrara, Don Garnett, Ed Jenkins, Chip Kobulnicky, Tom Jones,
B.-I. Jun, Jay Lockman, Blair Savage, Bart Wakker and the unknown
referee of a previous version of this paper. We are grateful to Gary
Ferland for the use of the photoionization code CLOUDY, and RAB would
like to thank Evan Skillman for his forbearance in allowing the time
to complete this work. A special debt of thanks goes Ron Reynolds,
whose work on the warm ionized medium was crucial in developing this
paper.  Finally, we would like to thank the referee, Lyman Spitzer,
Jr., for useful criticism and encouragement. This work was supported
by NASA LTSARP grant NAGW-3189 and the Minnesota Supercomputer
Institute.

\appendix
\section{Convergence to Terminal Velocity in a Uniform and Porous
Medium}

In a uniform density medium with constant gravitational acceleration,
a falling cloud will ultimately reach a terminal velocity. It is convenient
therefore to have a formulation that indicates how quickly a cloud
approaches terminal velocity, a value that can be compared with how
quickly the local terminal velocity changes.

Consider the case of a cloud in a uniform density gas with constant
$C_{D}$ and $g$, writing equation (1) in dimensionless form where
$\tilde{v}=v/v_{T}$, $\tilde{t}=t/t_{T}$, and $\tilde{z}=z/z_{T}$,
where $v_{T}$ is a positive definite quantity, and negative velocities
indicate motion toward the Galactic plane. The ``convergence time'',
$t_{T}=v_{T}/g$, and ``convergence length'', $z_{T}=v_{T}^{2}/g$ are
the typical time and length scales for a cloud to approach the
terminal velocity. The equation of motion is

\begin{equation} \label{eq-momsc}
\frac{d\tilde{v}}{d\tilde{t}}=\tilde{v}^{2}-1~.
\end{equation}

Leaving aside the case of cloud initially moving away from the
Galactic plane (in which case gravitational deceleration and drag have
the same sign), the solution for equation (A1)
breaks into the case in which the cloud is
accelerating to the terminal speed from below ($0> \tilde{v}_{o} >
-1$) and the case in which the cloud is decelerating
to the terminal speed ($-1> \tilde{v}_{o} > -\infty$).

Solving equation (A1) for the velocity as a function of time yields

\begin{equation}
\tilde{v}=\frac{\kappa \pm e^{2 \tilde{t}}}{\kappa \mp e^{2 \tilde{t}}} ~,
\end{equation}

where

\begin{equation}
\kappa = \pm \frac{\tilde{v}_{o}+1}{\tilde{v}_{o}-1} ~,
\end{equation}

where the upper sign is for the decelerating case, and
the lower is for the accelerating case. The position as a function of
time is

\begin{equation}
\tilde{z}-\tilde{z_{o}}=\ln (\frac{\kappa \mp 1}{\kappa \mp e^{2\tilde{t}}}) + 
\tilde{t}~.
\end{equation}

The relationship between position and velocity for both cases is

\begin{equation}
\tilde{z}-\tilde{z_{o}}=\frac{1}{2} \ln \left[ \frac{\tilde{v}^{2}-1}
{\tilde{v}_{o}^{2}-1} \right]~.
\end{equation}

How quickly does the cloud approches its terminal
speed? The time at which the cloud has reached a velocity midway
between its original velocity and the terminal velocity is 
$\tilde{v_{1/2}}=(\tilde{v_{o}}-1)/2$. For both the accelerating and
decelerating cases this is 

\begin{equation}
\tilde{t}_{1/2}=\frac{1}{2}\ln ( \frac{\tilde{v_{o}}-3}{\tilde{v_{o}}-1} )=c_{t}~.
\end{equation}

For $v_{o}$ between 0 and -2,  $c_{t}$ ranges from 0.55 to 0.25. This corresponds to a distance travelled of 

\begin{equation}
\tilde{z}_{1/2}=\ln (\frac{1}{2}\sqrt{ \frac{\tilde{v_{o}}-3}{\tilde{v_{o}}-1}} )=c_{z}~,
\end{equation}

where $c_{z}$ ranges from 0.14 to 0.43 for $v_{o}$ between 0 and
-2. Therefore, the cloud approaches the terminal velocity in
approximately one-quarter of a convergence length, $z_{T}$. In
dimensional coordinates then the typical lengthscale for the cloud to
approach the terminal velocity is 

\begin{equation}
l_{v} \cong 65~pc (\frac{c_{z}}{0.25})(\frac{N_{H~I}}{10^{19}~cm^{-
2}})(C_{D}f_{c})^{-1}(\frac{n_{H,o}}{0.025~cm^{-3}})^{-1}e^{z/H}~.
\end{equation}

By comparing this value with the terminal velocity curves in
Figure 2, one can see that clouds with $z ~^{<}_{\sim}
1~{\rm kpc}$ and $N(H~I) ~^{<}_{\sim} 10^{19}~{\rm cm^{-2}}$, will tend to
converge to the terminal velocity.

How does patchiness of the interstellar medium affect this convergence
to a terminal velocity?  The patchiness is characterized by two
parameters: $f$, the (linear) filling fraction of the dense gas, and
$\tilde{L}$, the normalized cell length which contains one filled
region and one unfilled region.
The density in the occupied zone is  $n_{h}/f$; 
equation (A1) must be modified accordingly.  

The cloud velocity oscillates around some mean value with a range
which depends upon the value of $f$ and $\tilde{L}$. It is possible to
solve for the mean value and range analytically.  The cloud alternates
between free-fall and deceleration. It eventually reaches a
steady-state such that the path lengths, $\Delta \tilde{z}$, of each
phase are equal. During the free-fall phase,

\begin{equation}
\Delta \tilde{z}=(1-f)\tilde{L}=\frac{1}{2}(\tilde{v}_{f}^{2}-\tilde{v}_{s}^{2})~,
\end{equation}

where $\tilde{v}_{f}^{2}$ is the velocity at the fast end of the oscillation
and  $\tilde{v}_{s}^{2}$ is the slow end. The path length covered during
the deceleration is

\begin{equation}
\Delta \tilde{z}=f \tilde{L}=\frac{1}{2} f \ln \left[\frac{\tilde{v}_{f}^{2}-
f}{\tilde{v}_{s}^{2}-f} \right]~.
\end{equation}

Setting these two pathlengths equal, the upper and lower
end of the velocity range are

\begin{equation}
|\tilde{v}_{s}|=\sqrt{2\tilde{L}(1-f)(e^{2\tilde{L}}-1)^{-1} + f}~,
\end{equation}

and

\begin{equation}
|\tilde{v}_{f}|=\sqrt{2\tilde{L}(1-f)e^{2\tilde{L}}(e^{2\tilde{L}}-1)^{-1} + f}~.
\end{equation}

The resulting HWZI, $\Delta
\tilde{v}=\frac{1}{2}|\tilde{v}_{f}-\tilde{v}_{s}|$ is plotted in Figure A1. The results separate
into two cases, one for cell lengths that are less than the
convergence length, $z_{T}$, and one for cell lengths that are
greater.

For $\tilde{L} \leq 1$, the length scale of the inhomogeneity is
smaller than the distance the cloud needs to travel to reequilibrate
its velocity. It then feels a drag force which on average is what the
cloud would feel in the uniform density medium, and its average velocity 
is the same as the uniform density case. However, its velocity will
oscillate around this average value where the oscillation depends
on the filling factor. At $\tilde{L}=1$, the cloud velocity is on average
$<\tilde{v}>=-1$, and lies in the range $<\tilde{v}> \pm \Delta
\tilde{v}$, where $\Delta \tilde{v}=\case{1}{2}(1-f)$.  As the filling factor
increases, $f \rightarrow 1$, and $\Delta \tilde{v} \rightarrow
0$. Similarly, as $\tilde{L} \rightarrow 0$, $\Delta \tilde{v}
\rightarrow 0$.

For larger inhomogeneities, $\tilde{L} \geq 1$, the mean speed exceeds
$1$. In the filled region, the cloud will have time to approach the
local terminal velocity. This value, which is the lower bound on the
cloud speed, is  $\tilde{v}_{s}=-\sqrt{f}$.  When the cloud
reaches the unfilled region, it will fall ballistically until
$\tilde{v}_{f}=-\sqrt{2 \tilde{L}(1-f)+f}$. For large $\tilde{L}$ or
small $f$, this means that the upper velocity bound will increase
resulting in an increase in both the mean velocity and $\Delta
\tilde{v}$.  \S~\ref{sec-dstruct} discusses how observational
limits on $\Delta \tilde{v}$ can be used to estimate the porosity of
the ISM at high $z$.

\vfill
\eject

\vfill
\eject

\begin{figure}
\figurenum{1}
\plotone{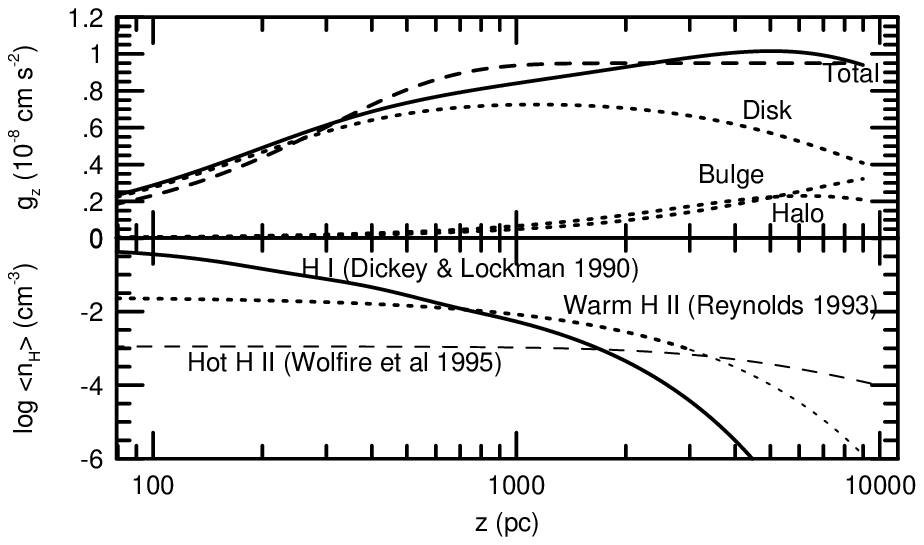}
\caption{Input parameters for terminal velocity model. [Top panel] Gravitational acceleration ($cm~s^{-2}$) in z direction from
Wolfire et al. (1995).  The total acceleration is shown with a solid
line; the contribution to this total from the disk, bulge, and halo
are indicated with dotted lines. The dashed line indicates a fit to
the potential,  $g(z)=9.5 \times 10^{-9} \tanh(z/400~pc)$. [Bottom panel]
Vertical density structure of the gaseous disk and halo. The
mean hydrogen particle densities of neutral H~I (Dickey \& Lockman 1990),
the warm ionized H II (Reynolds 1993), and a possible hot halo (Wolfire
et al. 1995) are indicated.
}
\end{figure}\label{fig-grav}

\vfill\eject

\begin{figure}
\figurenum{2}
\plotone{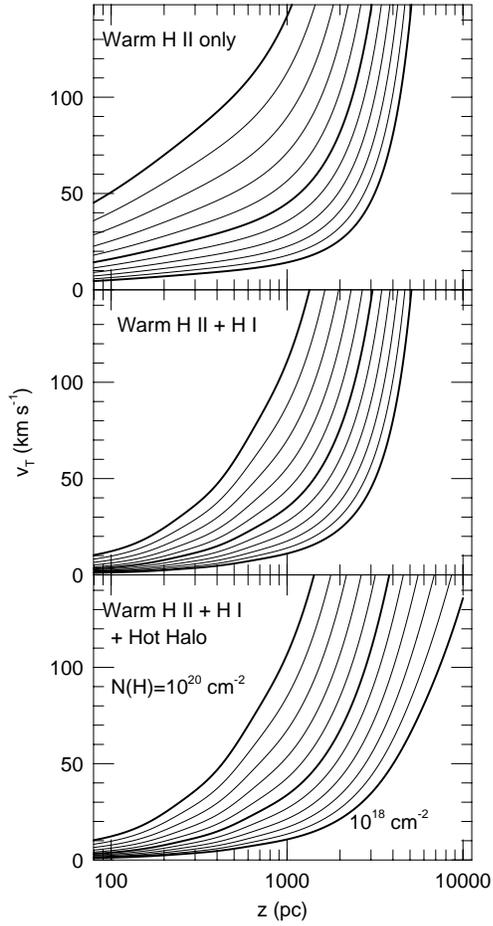}
\caption{Terminal velocity as a function of height $z$ and total column
density $N(H)$, assuming $C_{D}=1$, for Warm H~II only
[top panel], Warm H~II plus H~I layer [middle panel], and
Warm H~II + H~I layer + Hot halo [bottom panel].  Curves range from 
$N(H)=10^{18}~{\rm cm^{-2}}$ to $10^{20}~{\rm cm^{-2}}$, separated by 0.2 dex.
Given a cloud velocity and total column density, this figure can be used
to predict the cloud distance for a given halo density model.}
\end{figure} \label{fig-vt}

\vfill\eject

\begin{figure}
\figurenum{3}
\plotone{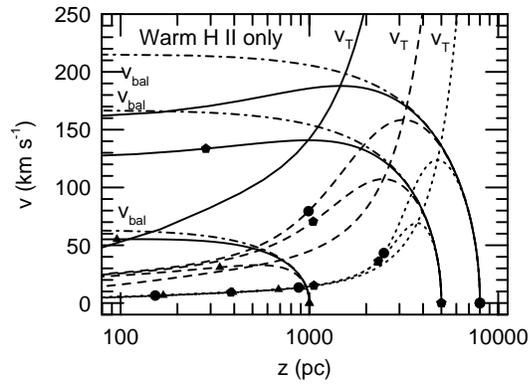}
\caption{Trajectories of clouds dropped at rest from heights of
$z_{i}=$ 1 (triangle), 5 (pentagon), 8 (circle) ${\rm kpc}$ in a
gaseous halo with density model A. Cases are shown for clouds with
$N_{H~I}=10^{18}~{\rm cm^{-2}}$(dotted) , $10^{19}~{\rm cm^{-2}}$(dashed), and
$10^{20}~{\rm cm^{-2}}$(solid).  Ballistic trajectories from each starting
point (dot-dash) and the local terminal velocity curves for each
column density (with the same line type as above) are indicated. The
points note the location after $n$ ($n=1,2,..$) free-fall times.}
\end{figure} \label{fig-traj}

\vfill\eject

\begin{figure}
\figurenum{4}
\plotone{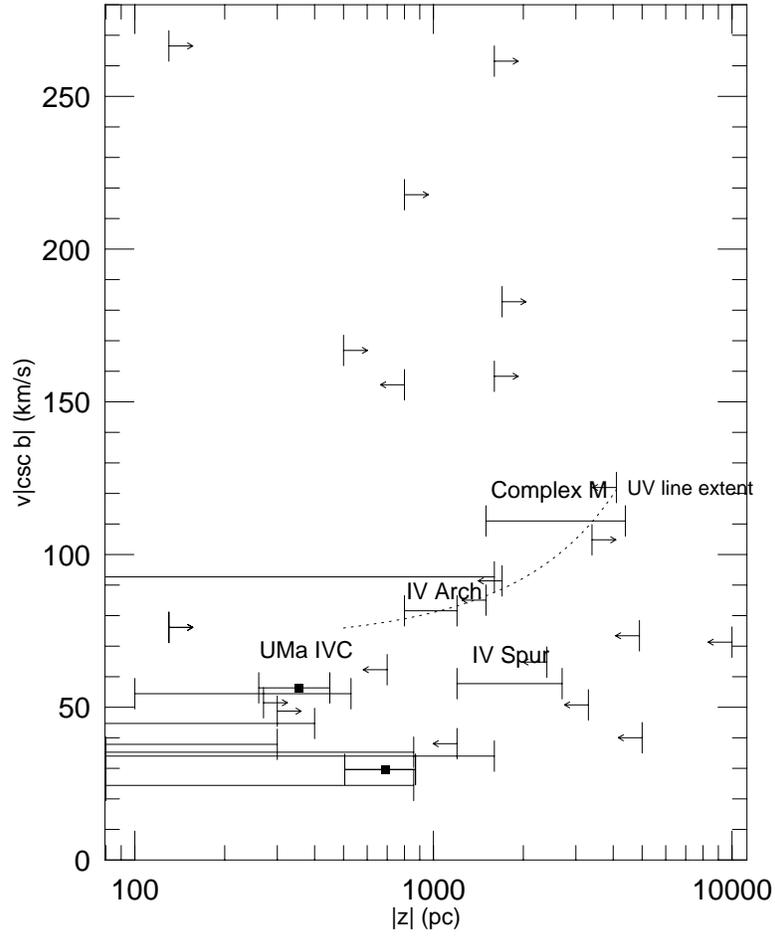}
\caption{Velocity-distance relation for infalling halo clouds.  $|v~
{\rm csc}~ b|$ is plotted for halo clouds with $|z| ^{>}_{\sim}
200~pc$, $v < 0$, and $|b|>30^{\circ}$. The dotted curve is a
fit to the maximum negative velocity extent of UV absorption lines in
Figure 4 of Danly (1989). Note the trend of cloud velocity increasing
with $z$.}
\end{figure} \label{fig-clouddata}

\vfill\eject

\begin{figure}
\figurenum{5}
\plotone{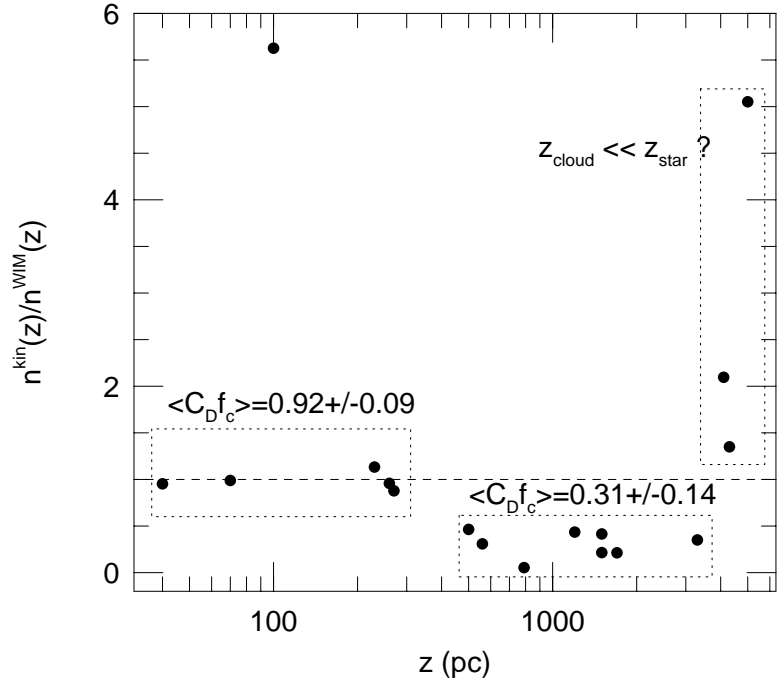}
\caption{A comparison of two independent methods to estimate the vertical 
density structure of the ISM. Pulsar dispersion measures are used
to obtain $n^{WIM}(z)$ (Reynolds 1993), while the terminal velocity model is used
to obtain $n^{kin}(z)$. The ratio, which is equivalent to deriving  $(C_{D}f_{c})$ for a given density model, is plotted as a function of $z$. If it
assumed that $C_{D}\approx 1$, the trend is most simply explained by
a decrease in the cloud ionization fraction at $z \approx 400~pc$, and
a top to the cloud layer at about $z \approx 3.3 kpc$. }
\end{figure}

\vfill\eject

\begin{figure}
\figurenum{A1}
\plotone{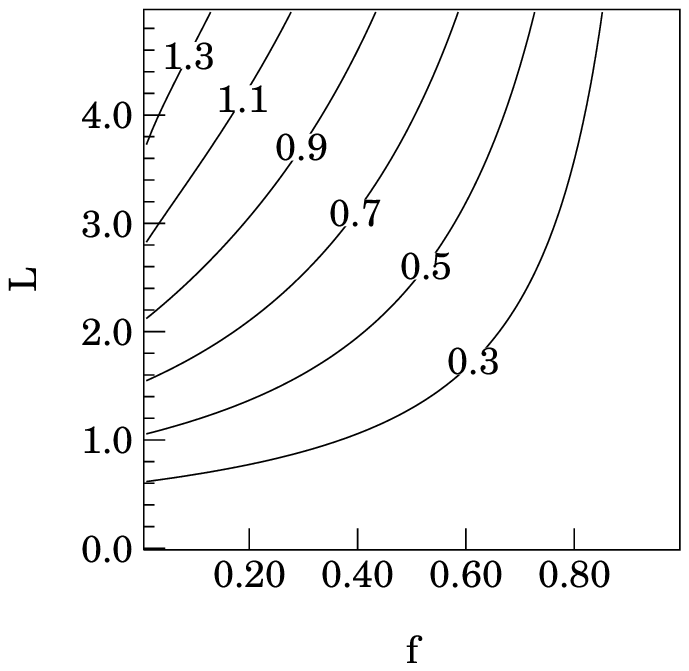}
\caption{Contours of velocity range, $\Delta \tilde{v}=\frac{1}{2}(|\tilde{v}_{f}-\tilde{v}_{s}|)$  for a cloud in a 
patchy medium characterized by filling factor, $f$, and cell length,
$\tilde{L}$.}
\end{figure} \label{fig-dv}

\vfill\eject

\begin{planotable} {lrccccc} 
\tablecaption{Terminal Velocity Model Predictions for Clouds of Known Distance
\tablenotemark{a}} 
\tablehead{ \colhead{Cloud} & \colhead{b} & \colhead{$v$} & \colhead{$N(H I)$}
  &\multicolumn{2}{c}{Predicted $z~(kpc)$} & Observed $z$\tablenotemark{b} \\ 
\colhead{} & \colhead{} &
\colhead{$km~s^{-1}$}   & \colhead{$\log~cm^{-2}$}  & 
\colhead{A}   & \colhead{B} &
 \colhead{(kpc)} } 
\startdata
Complex M & 62 &  $-98 \pm 1$\tablenotemark{c} & $18.6 \pm 0.1$\tablenotemark{d} &  $3.44 \pm 0.02$ & 
$3.46 \pm 0.02$ & $1.5-4.1$\tablenotemark{d} \nl
$G 259+61-64$\tablenotemark{j} & 61 & -64.2 & 18.36 & 3.22 & 3.24 & $<4.90$ \nl
$G 22+52-43$\tablenotemark{j} & 52 & -42.9 & 18.91 & 1.69 & 1.84 & $0.10-0.53:$ \nl
$G 218-63-30$\tablenotemark{j} & -63 & -30.3 & 18.45 & 1.68 & 1.84 & $0.08-1.60:$ \nl
IV Arch & 62  & $-66.3 \pm 0.3$\tablenotemark{e} & $18.85 \pm 0.03$\tablenotemark{e,f} &  $1.56 \pm 0.01$ &
 $1.74 \pm 0.01$ & $0.78-1.5$\tablenotemark{g} \nl
$G 322+55-29$\tablenotemark{j} & 55 & -28.9 & 18.76 & 1.22 & 1.46 & $0.08-0.86$ \nl  
$G 113+50-71$\tablenotemark{j} & 50 & -70.9 & 19.77 & 0.96 & 1.26 & $0.06-1.60$ \nl
$G 68+64-22$\tablenotemark{j} & 64 & -21.5 & 19.24 & 0.15 & 0.51 & $0.10-0.79$ \nl
UMa IVC & 53 & $-44.9 \pm 0.7$\tablenotemark{h} & $20.2 \pm 0.2$\tablenotemark{i} &  $0.10 \pm 0.02$ & 
$0.43 \pm 0.02$ & $0.29 \pm 0.08$\tablenotemark{h} \nl
$G 322+55-20$\tablenotemark{j} & 55 & -20 & 19.5 & 0.08 & 0.41 & $0.08-0.86$ \nl
\enddata
\tablenotetext{a}{Predictions assume $f_{c}=1$, except for IV Arch with $f_{c}=0.44$. Lowering $f_{c}$ will reduce distance estimate by $\Delta z \cong 910~pc(\ln{f_{c}^{-1}})$}
\tablenotetext{b}{Uncertain detection denoted by ``:''}
\tablenotetext{c}{Danly, Lee, \& Benjamin (1997)}
\tablenotetext{d}{Danly, Kuntz, \& Albert (1993)}
\tablenotetext{e}{Spitzer \& Fitzpatrick (1993)}
\tablenotetext{f}{N(H~I) estimated by using N(S II); see text for details.}
\tablenotetext{g}{Kuntz \& Danly (1996)}
\tablenotetext{h}{Benjamin et al (1996)}
\tablenotetext{i}{Snowden et al (1994)}
\tablenotetext{j}{All data from Albert et al (1993)}
\label{tbl-predict1} 
\end{planotable} 

\begin{planotable} {llllllll} 
\tablecaption{Terminal Velocity Model Predictions for Absorption Components Seen in HD 93521 ($z=1.5 \pm 0.4~kpc$; $b=62$)} 
\tablehead{ \colhead{Component} & \colhead{$v$} & \colhead{$N(S II)$}
& \colhead{$N(S III)$} & \colhead{$N(H I)_{s}$} & \colhead{$f_{c}$} 
& \multicolumn{2}{c}{Predicted $z~(kpc)$}  \\ 
\colhead{} & \colhead{$km~s^{-1}$}   & \colhead{$\log~cm^{-2}$}  & \colhead{$\log~cm^{-2}$} & 
\colhead{$\log~cm^{-2}$} & \colhead{} &\colhead{A}   & \colhead{B}  } 
\startdata
1 & $-66.3 \pm 0.3$ & $14.09 \pm 0.03$ & $13.05 \pm 0.21$ & $18.85 \pm 0.03$ & 0.44 & 
$1.56 \pm 0.01$  & $1.74 \pm 0.01$  \nl
2 & $-57.8 \pm 0.3$ & $14.50 \pm 0.02$ & $13.11 \pm 0.22$ & $19.26 \pm 0.02$ & 0.76 &
$1.00 \pm 0.01$  & $1.30 \pm 0.01$ \nl
3 & $-51.2 \pm 0.7$ & $14.53 \pm 0.03$ & $13.71 \pm 0.07$ & $19.29 \pm 0.03$ & 0.43 &
$0.39 \pm 0.02$  & $0.80 \pm 0.02$  \nl
4 & $-38.8 \pm 0.5$ & $14.02 \pm 0.07$ & $13.14 \pm 0.31$ & $18.78 \pm 0.07$ & 0.35 &
$0.64 \pm 0.02$  & $1.02 \pm 0.02$  \nl
5 & $-29.1 \pm 0.5$ & $13.16 \pm 0.20$ &                  & $17.92 \pm 0.20$ & 1.0  &
$2.68 \pm 0.03$  & $2.74 \pm 0.03$ \nl
6 & $-18.2 \pm 0.4$ & $14.51 \pm 0.02$ &                  & $19.27 \pm 0.02$ & 1.0  &
$0.11 \pm 0.04$  & $0.44 \pm 0.04$ \nl
7 & $-10.2 \pm 0.3$ & $14.66 \pm 0.02$ &                  & $19.42 \pm 0.02$ & 1.0  &
$0.03 \pm 0.05$  & $0.21 \pm 0.05$ \nl
\enddata
\label{tbl-predict2} 
\end{planotable} 

\begin{planotable} {lrrrllll} 
\tablecaption{Kinematically Derived Halo Density and Drag coefficient\tablenotemark{a}} 
\tablehead{ \colhead{Star} & \colhead{b} & \colhead{$z$} & \colhead{$v_{-}$} &
\colhead{$N(H I)$\tablenotemark{b}} & \colhead{$n^{kin}_{h}(z)C_{D}f_{c}$} &
 \multicolumn{2}{c}{$C_{D}f_{c}$\tablenotemark{c}} \\ 
\colhead{} & \colhead{} & \colhead{kpc} &
\colhead{$km~s^{-1}$}   & \colhead{$\log~cm^{-2}$}  &
\colhead{$(cm^{-3})$} & \colhead{A}   & \colhead{B} 
 } 
\startdata
HD 100340   &   $61$  &   $5.00$  &     $-68$  &   18.25 &  $5.19 \times 10^{-4}$  &  5.052  &  5.039 \nl
HD 121968   &   $56$  &   $4.30$  &     $-57$  &   17.93 &  $2.99 \times 10^{-4}$  &  1.350  &  1.341 \nl
BD $+38~2182$\tablenotemark{d} &   $62$  &   $4.10$  &     $-98$  &   18.60 &  $5.79 \times 10^{-4}$  &  2.096  &  2.078 \nl
HZ 25       &   $80$  &   $3.30$  &     $-84$  &   17.94 &  $2.34 \times 10^{-4}$  &  0.351  &  0.342 \nl
HD 121800   &  $ 50$  &   $1.70$  &     $-91$  &   18.92 &  $8.24 \times 10^{-4}$  &  0.213  &  0.172 \nl
HD 93521\tablenotemark{d}    &  $ 62$  &   $1.50$  &     $-66$  &   18.85 &  $1.99 \times 10^{-3}$  &  0.415  &  0.314 \nl
HD 219188   &  $-50$  &  $-1.50$  &     $-48$  &   18.47 &  $1.03 \times 10^{-3}$  &  0.215  &  0.163 \nl
HD 97991    &  $ 52$  &   $1.20$  &     $-49$  &   18.92 &  $2.91 \times 10^{-3}$  &  0.435  &  0.292 \nl
HD 91316    &  $ 53$  &   $0.79$  &     $-56$  &   18.33 &  $5.66 \times 10^{-4}$  &  0.054  &  0.029 \nl
HD 220172   &  $-63$  &  $-0.56$  &     $-35$  &   18.68 &  $4.18 \times 10^{-3}$  &  0.309  &  0.122 \nl
HD 137569   &  $ 52$  &   $0.50$  &     $-35$  &   19.06 &  $6.69 \times 10^{-3}$  &  0.464  &  0.159 \nl
HD 87015    &  $ 51$  &   $0.27$  &     $-30$  &   19.43 &  $1.63 \times 10^{-2}$  &  0.878  &  0.151 \nl
HD 113001\tablenotemark{e} & $81$  &  $0.26$   &     $-35$  &   19.30 &  $1.80 \times 10^{-2}$  &  0.958  &  0.159 \nl
HD 100600   &  $ 69$  &  $0.23$   &     $-37$  &   19.54 &  $2.20 \times 10^{-2}$  &  1.133  &  0.164 \nl
HD 219688   &  $-62$  & $-0.10$   &     $-13$  &   19.73 &  $1.26 \times 10^{0}$   &  5.627  &  0.330 \nl
HD 138749\tablenotemark{e} & $55$  &  $0.07$   &     $-28$  &   19.90 &  $2.29 \times 10^{-2}$  &  0.989  &  0.049 \nl
HD 120315\tablenotemark{e} & $65$  &  $0.04$   &     $-11$  &   19.18 &  $2.28 \times 10^{-2}$  &  0.953  &  0.042 \nl
\enddata
\tablenotetext{a}{Absorption velocity, $v_{-}$ and star distance, $z$, from Danly(1989) unless otherwise noted.}
\tablenotetext{b}{N(H I) calculated by integrating 21 cm emission profiles of Danly et al (1992) over velocity range
$\Delta v=v_{-} \pm 5 {\rm km~ s^{-1}}$, unless otherwise noted.}
\tablenotetext{c}{$n^{kin}_{h}(z)$ divided by density structure of model A or B}
\tablenotetext{d}{Data updated since Danly (1989). See references for Complex M (BD+38 2182) and
IV Arch (HD 93521) in Table 1.}
\tablenotetext{e}{N(H~I) estimated from 21 cm emission maps of Kuntz \& Danly (1996)}
\label{tbl-halostruct} 
\end{planotable} 

\end{document}